\documentclass[10pt]{article}

\usepackage[utf8]{inputenc}
\usepackage{graphicx}
\usepackage{caption}
\usepackage{amsmath} 
\usepackage[squaren, Gray, cdot]{SIunits}
\usepackage{hyperref}

\begin{document}
\title{High frequency optomechanical disk resonators in III-V ternary semiconductors}
\date{\vspace{-8ex}}
\maketitle

\author{
\begin{center}
\textbf{Biswarup Guha, Silvia Mariani, Giuseppe Leo, Ivan Favero$^*$}
\\ 
\textsl{Mat\'{e}riaux et Ph\'{e}nom\`{e}nes Quantiques, Universit\'{e} Paris Diderot \\ CNRS UMR 7162, Sorbonne Paris Cit\'{e} \\ 10 rue Alice Domon et L\'{e}onie Duquet, 75013 Paris, France}
\end{center}
}
\author{
\begin{center}
\textbf{Aristide Lema\^{i}tre} 
\\
\textsl{Centre de Nanosciences et de Nanotechnologies, CNRS \\ Université Paris Sud, Universit\'{e} Paris-Saclay C2N-Marcoussis \\ Route de Nozay, 91460 Marcoussis, France}
\end{center}
}

$^*$\textsl{ivan.favero@univ-paris-diderot.fr}

\begin{abstract}
Optomechanical systems based on nanophotonics are advancing the field of precision motion measurement, quantum control and nanomechanical sensing. In this context III-V semiconductors offer original assets like the heteroepitaxial growth of optimized metamaterials for photon/phonon interactions. GaAs has already demonstrated high performances in optomechanics but suffers from two photon absorption (TPA) at the telecom wavelength, which can limit the cooperativity. Here, we investigate TPA-free III-V semiconductor materials for optomechanics applications: GaAs lattice-matched In$_{0.5}$Ga$_{0.5}$P and Al$_{0.4}$Ga$_{0.6}$As. We report on the fabrication and optical characterization of high frequency (500-700 MHz) optomechanical disks made out of these two materials, demonstrating high optical and mechanical Q in ambient conditions. Finally we achieve operating these new devices as laser-sustained optomechanical self-oscillators, and draw a first comparative study with existing GaAs systems.
\end{abstract}

\section{Introduction}

The interaction between light and mechanical motion is at the core of a field of research, optomechanics, which has been rapidly growing lately \cite{Favero2009, Aspelmeyer2014}. Optomechanical cavities are employed to detect mechanical vibration \cite{Ding2010, Stapfner2009}, sense mass \cite{Liu2013, Yu2016, Santos2015} and study the quantum behavior of mechanical systems \cite{Teufel2011}. Diverse miniature optomechanical systems were developed out of different dielectric materials such as silicon (Si) \cite{Eichenfield2009,Metzger2008}, silica (SiO$_2$) \cite{Liu2013, Yu2016}, silicon nitride (SiN) \cite{Baker2012, Wiederhecker2009} or gallium arsenide (GaAs) \cite{Ding2010}. GaAs presents a series of assets for optomechanics applications: a strong photoelasticity \cite{Baker2014}, a surface dissipation that can be controlled by proper treatments \cite{Guha2017}, and a lattice matching with the family of aluminium and indium-rich GaAs ternary compounds. The latter fact permits the production of heteroepitaxial materials with designed photon/phonon interactions, based for example on quantum wells \cite{Jusserand2015}. Together with Si, GaAs is probably amongst the most mature materials in semiconductor technology but both suffer from two photon absorption (TPA) at the telecom wavelength \cite{Johnson2006, Parrain2015}. At large optical power, such multi-photon absorption processes dominate optical dissipation and degrade the attainable cooperativity in the widely used linearized regime of optomechanics. 

In this work, we investigate alternative III-V semiconductors that possess optomechanical assets similar to those of GaAs: the ternary compounds indium gallium phosphide (In$_{0.5}$Ga$_{0.5}$P) and aluminium gallium arsenide (Al$_{0.4}$Ga$_{0.6}$As). Unlike Si and GaAs, both InGaP and AlGaAs are free of TPA at the telecom wavelength. InGaP photonic crystal nanophotonic cavities were reported to exhibit an optical quality factor Q$_{\text{opt}}$ of a million \cite{Combrie2009}, while AlGaAs whispering gallery optical resonators were shown to permit enhanced second harmonic generation \cite{Mariani2014article}. The mechanical and optomechanical properties of both materials are however little known \cite{Cole2014}. Here we report on the fabrication and investigation of both InGaP and AlGaAs high frequency ($>100$ MHz) optomechanical disk resonators. A complete set of optomechanical measurements on both families of resonators is presented along with a study of their dynamical behaviour under illumination.

\section{Fabrication and spectroscopy methods}

InGaP disk resonators are fabricated out of an extra-pure wafer consisting of a layer of In$_{0.5}$Ga$_{0.5}$P (\( 200\ \)nm) and of GaAs (\( 1.6\ \mu \)m thick) grown by Metal-Organic Chemical Vapour Deposition (MOCVD) on a semi-insulating GaAs substrate. The disks are patterned on the surface by e-beam lithography using a negative resist (MaN 2403) and fabricated using subsequent etching steps. A first non-selective wet-etching is carried out at 4$^{\circ}$C in a solution consisting of hydrobromic acid, potassium dichromate and acetic acid in equal proportion (BCK solution), giving rise to InGaP/GaAs pillar structures. The pedestals of the disks are obtained by selective under-etching of GaAs in a solution of citric acid and hydrogen peroxide, whose endothermic reaction is sustained by heating the solution to 30$^{\circ}$C. Figure 1(a) shows a Scanning Electron Microscope (SEM) image of a resulting InGaP disk atop a GaAs pedestal. 
\begin{figure}[htbp]
\centering\includegraphics[width=\textwidth]{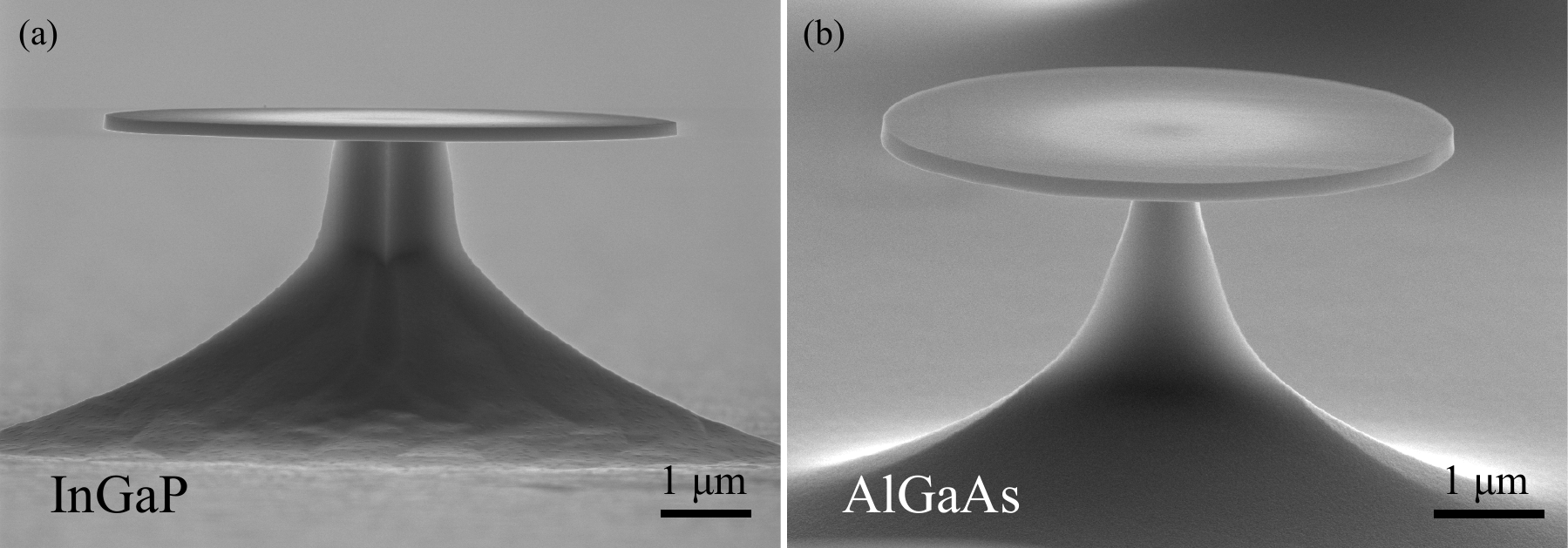}
\captionsetup{labelsep=period}
\caption{An (a) In$_{0.5}$Ga$_{0.5}$P and (b) Al$_{0.4}$Ga$_{0.6}$As optomechanical disk over a GaAs pedestal.}
\end{figure}

The AlGaAs wafer consists of a top Al$_{0.4}$Ga$_{0.6}$As layer (\( 150\ \)nm) over a sacrificial GaAs layer (\( 1.5\ \mu \)m), grown by molecular beam epitaxy on a semi-insulating GaAs substrate. The fabrication steps for AlGaAs disks are similar to InGaP, except for the selective under-etching of GaAs, which is carried in a solution of H$_2$O$_2$ and NH$_4$OH \cite{Mariani2013, Mariani2014}. Figure 1(b) shows an AlGaAs disk on a GaAs pedestal. 

Both resonators are optically addressed by fiber-taper evanescent coupling techniques \cite{Ding2011, Ding2014, Ding2010SPIE}. In order to prevent coupling of light from the fiber taper to the sample substrate, the disks are elevated over the substrate on a mesa structure. The latter is patterned by photolithography (resist S1828) and wet etched in BCK at room temperature. Figure 4 of the Appendix shows disk resonators on such mesa.

\section{Optical and mechanical measurements}

A disk structure is both an optical and mechanical resonator \cite{Ding2014}, supporting optical Whispering Gallery Modes (WGMs) together with mechanical Radial Breathing Modes (RBMs) of a contour type. Here we perform optical spectroscopy of these modes. When a fiber taper is evanescently coupled to a disk, WGMs appear as dips in the fiber optical transmission, as shown in Fig. 2. Figure 2(a) is the optical spectrum of an InGaP disk measured in the wavelength range \( 1500-1600\ \)nm, in the under-coupled regime for all modes. Several WGM resonances appear in the spectrum with a high contrast, and correspond to TE-polarized (in plane) modes of radial order $2\leq p\leq 4$. The loaded optical quality factor (Q$_{\text{opt}}$) evolves between $2\times 10^3$ and $6\times 10^4$, with the best intrinsic Q$_{\text{opt}}$ attaining $10^5$. While a complete study of optical dissipation \cite{Parrain2015} was not yet carried on our InGaP disks, our current understanding is that such Q$_{\text{opt}}$ may be increased by improving the fabrication procedure, as the material loss sits well below the here-observed level \cite{Combrie2009}. The drop of Q factor observed in Fig. 2 as the radial order p increases is for example compatible with a situation of optical scattering at the surfaces, as observed in Fig. 8 of \cite{Parrain2015}. 
\begin{figure}[h!]
\centering\includegraphics[width=\textwidth]{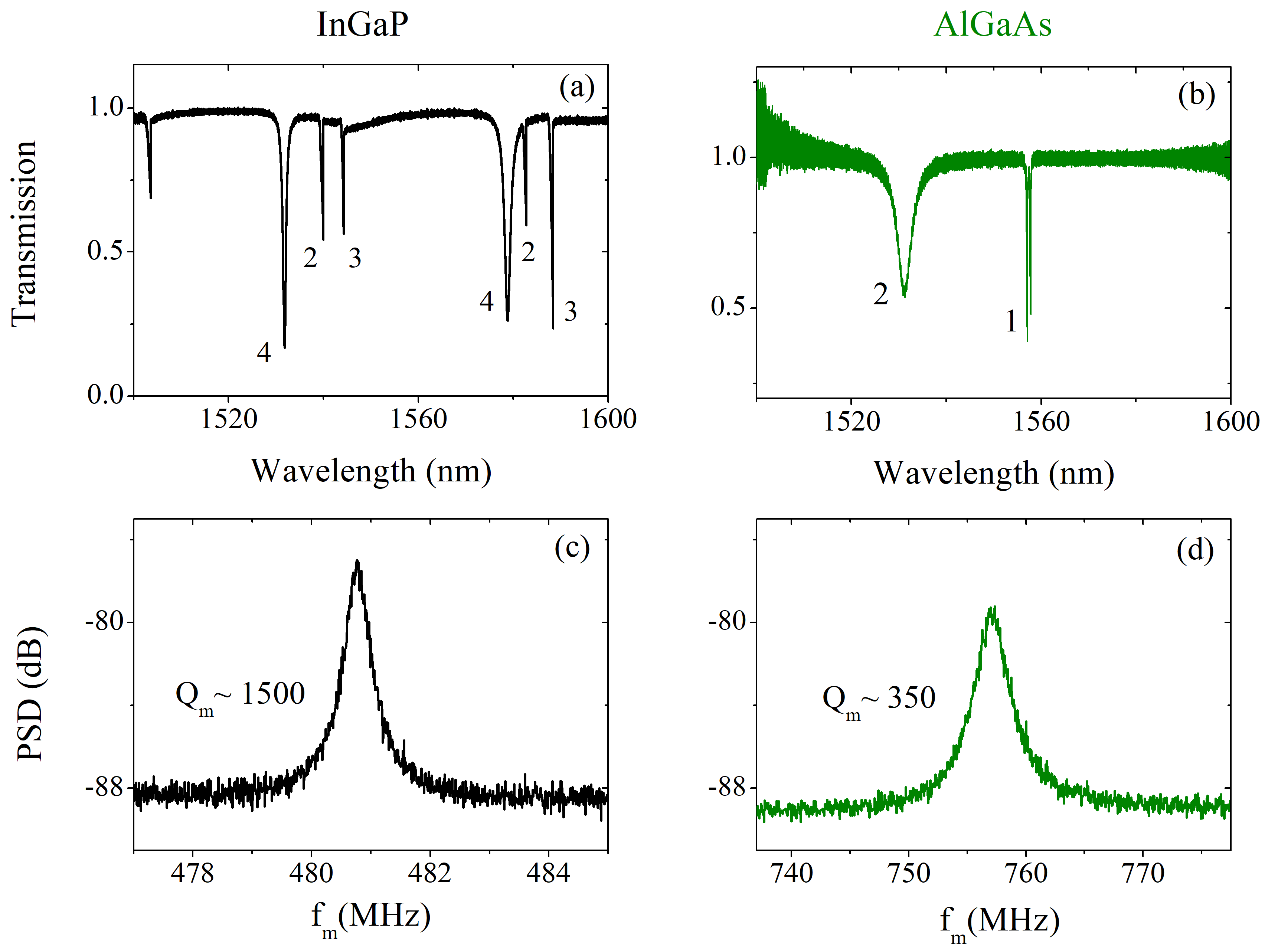}
\captionsetup{labelsep=period}
\caption{Optical spectrum of (a) an InGaP disk of radius \( 3.25\ \mu \)m and thickness \( 200\ \)nm and (b) an AlGaAs disk of radius \( 2\ \mu \)m and thickness \( 150\ \)nm. The spectra are acquired with an optical power of \( 50\ \mu \)W, measured at the output of the fiber. The radial order ($p$) of the WGMs is identified with the help of Finite Element Method (FEM) simulations and indicated. The probe light is TE-polarized (in the disk plane). (c) and (d) Mechanical spectra of the same disks as above, where the 1$^{\text{st}}$ order RBM is measured in the Brownian motion regime by optomechanical means.}
\end{figure}
The optical spectrum of an AlGaAs disk is presented in Fig. 2(b). TE WGMs of radial order $p = 1$ and 2 are measured with a loaded optical quality factor between $5\times 10^2$ and $1\times 10^4$, with the best intrinsic Q$_{\text{opt}}$ attaining $5\times 10^4$. Here again, a complete study of Q$_{\text{opt}}$ is beyond the scope of the present paper but the clear observation of a doublet structure in Fig. 2(b) is a marker of lack of symmetry of the disk structure. We also observed that Atomic Layer Deposition (ALD) of alumina at the surface of AlGaAs disks reduces heating effects, pointing towards the importance of residual surface absorption in these resonators. 

When tuning the laser wavelength on the flank of a WGM resonance, the mechanical motion of the disk resonator becomes imprinted onto optical intensity fluctuations at the output of the fiber, thanks to optomechanical coupling. The radio-frequency spectrum of the output light hence provides a mechanical spectrum. Figure 2(c) and (d) show the obtained resonances for the 1$^{\text{st}}$ order mechanical RBM of InGaP and AlGaAs disk resonators, measured in the Brownian motion regime in ambient conditions \cite{Ding2014}. These spectra were taken at the largest blue detuning accessible given our signal to noise ratio. According to our optomechanical modeling (see Appendix) back-action on the mechanical linewidth can be neglected under these conditions. The mechanical frequencies are respectively 480.77 MHz and 757.10 MHz, which are compatible within 5 $\%$ with FEM elastic simulations of the corresponding structures. The mechanical quality factor (Q$_{\text{m}}$) is $\sim$ 1500 for the InGaP disk and $\sim$ 350 for the AlGaAs disk, the difference between both resulting from a larger pedestal in the AlGaAs structure measured in Fig. 2, while air damping plays a secondary role \cite{Santos2015}. In the following, we investigate dynamical interactions between the here measured optical and mechanical modes of InGaP and AlGaAs disks. 

\section{Optomechanical self-oscillation}

For a laser blue detuned to the cavity and for sufficient optical power, the mechanical motion is amplified by optomechanical parametric gain. Figures 3(a) and (b) show the evolution of the optically measured mechanical spectrum of both the InGaP and AlGaAs disk as a function of normalized laser-cavity detuning $\Delta \omega /\kappa$ (see exact definition of this parameter in the Appendix). This evolution is measured from a large blue detuning to a smaller blue detuning.
\begin{figure}[htbp]
\centering\includegraphics[width=\textwidth]{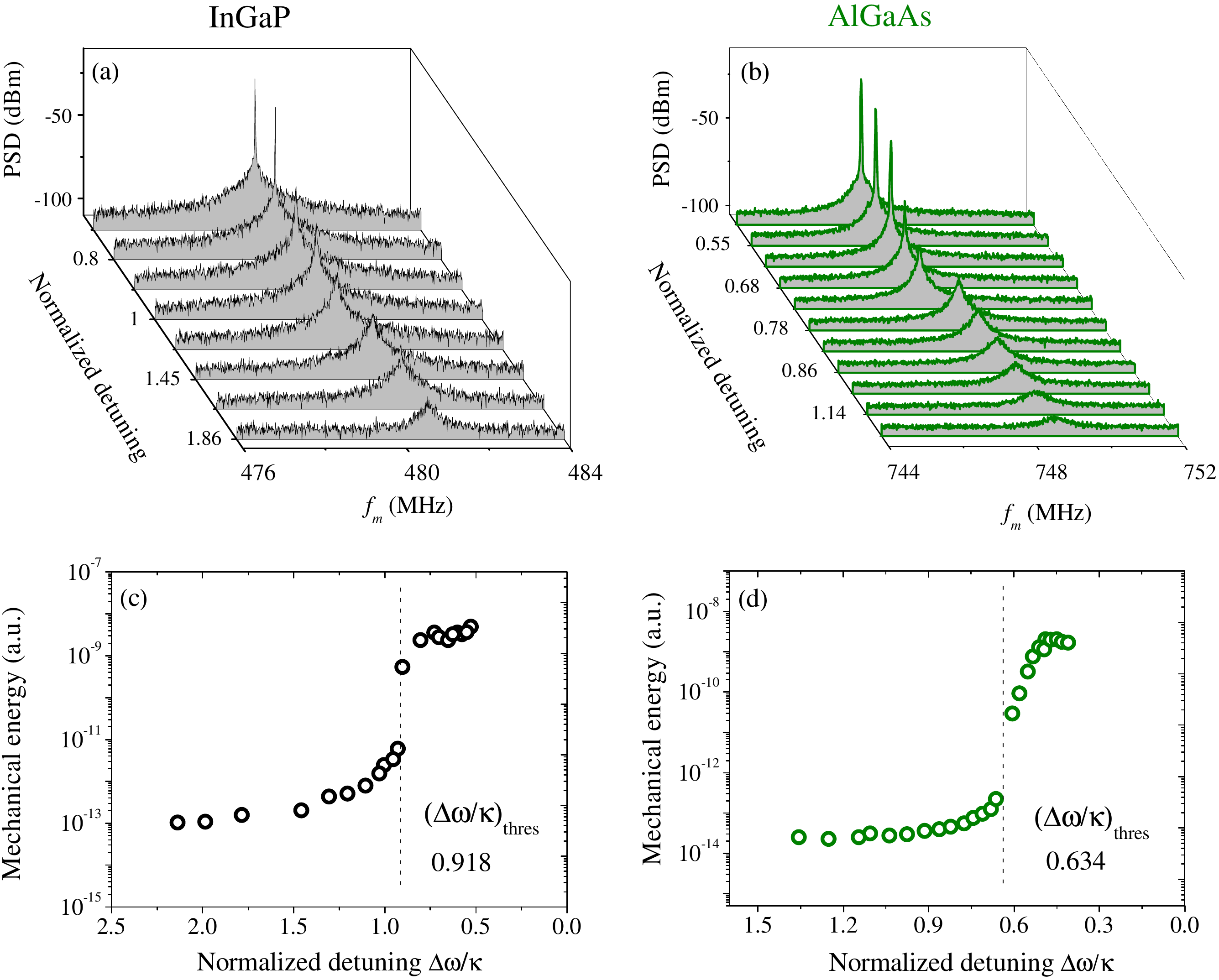}
\captionsetup{labelsep=period}
\caption{Optomechanical self-oscillation of InGaP and AlGaAs disk resonators. (a,b) Evolution of the mechanical spectrum as a function of normalized detuning $\Delta \omega /\kappa$. (c,d) Mechanical energy as a function of $\Delta \omega /\kappa$. The measurements of Fig. 3(a) are obtained for an optical power of \( 1.1\ \)mW in the fiber taper, using a TE WGM ($p=3$) with a loaded Q$_{\text{opt}}$ = $5\times 10^4$, while those of Fig. 3(b) are obtained for \( 4.2\ \)mW of optical power, using a TE WGM ($p=1$) with loaded Q$_{\text{opt}}$ = $7\times 10^3$.The dashed line shows the threshold of self-oscillation.}
\end{figure}

When the laser is tuned on the blue flank of a WGM resonance, the mechanical vibration is first amplified. Above a certain threshold, the optomechanical amplification counterbalances natural mechanical losses: at that stage, the mechanical motion becomes self-sustained and acquires a coherent harmonic trajectory \cite{Metzger2008, Carmon2005}. The onset of this self-oscillation regime corresponds to the line narrowing observed in Fig. 3(a) and (b). It is directly apparent in Fig. 3(c) and (d), where we show the mechanical energy (in arbitrary unit) as a function of the normalized detuning. The mechanical energy is proportional to the area under the curve in the mechanical spectrum. The self-oscillation threshold is marked by a dashed line, and it can be modelled as the point of cancellation of the effective mechanical damping $\Gamma_{m,eff}$ (the natural damping $\Gamma_m$ minus the optomechanical amplification). $\Gamma_{m,eff}$ is derived from the linearized equations of optomechanics including radiation pressure, electrostriction and photothermal forces (see Appendix). The force per photon associated to radiation pressure ($F_{rp}^1$) and electrostriction ($F_{el}^1$) are obtained through the relations $F_{rp}^1= \hbar g_{om}^{geo}$ and $F_{el}^1= \hbar g_{om}^{pe}$, after having computed by FEM the geometric (photoelastic) frequency-pull parameter $g_{om}^{geo}$ ($g_{om}^{pe}$) ($g_{om}=-\frac{\partial\omega_{cav}}{\partial x}$, differential shift of the cavity frequency $\omega_{cav}$ for an elementary mechanical displacement $\partial x$ \cite{Baker2014}). The photothermal force ($F_{pth}$) is associated to a thermal distortion of the mechanical structure, and is hence a consequence of optical absorption in the cavity. The latter is evaluated by fitting the thermo-optic wavelength drag of the WGM resonance at high optical power (see Fig. 5 of Appendix) \cite{Parrain2015, Carmon2004, Almeida2004, Weidner2007}, with as fit parameter the product $\kappa_{abs}\times R_{th}$, where $\kappa_{abs}$ is the rate of linear absorption of cavity photons and $R_{th}$ is the thermal resistance of the disk (see Appendix). The thermal resistance is simulated by FEM using the SEM-measured dimensions of the disk and pedestal, as well as the material thermal properties, such that $\kappa_{abs}$ can eventually be extracted from the fit.
\begin{table}[h!]
\centering
\captionsetup{font=bf}
\captionsetup{labelsep=period}
\caption{Radiation pressure, Electrostriction and Photothermal force per photon, together with the thermal relaxation time and vacuum optomechanical coupling $\boldsymbol{g_0}$ for the considered InGaP and AlGaAs disk resonator.}
\smallskip
\begin{tabular}{c  c  c  c  c  c}
 \hline
 \textbf{Disk}
 &
 $\boldsymbol{F_{rp}^1}$ \textbf{(N)}
 &
 $\boldsymbol{F_{el}^1}$ \textbf{(N)}
 &
 $\boldsymbol{F_{pth}^1}$ \textbf{(N)}
 &
 $\boldsymbol{\tau_{th}}$ \textbf{(}$\boldsymbol{\mu}$\textbf{s)}
 &
 $\boldsymbol{g_0}$ \textbf{(kHz)} \\
\hline
\textbf{InGaP} & $1.92\times 10^{-14}$ & $2.68\times 10^{-14}$ & $2.42\times 10^{-9}$ & 1.8 & 390 \\
\hline
\textbf{AlGaAs} & $2.49\times 10^{-14}$ & $3.42\times 10^{-14}$ & $3.11\times 10^{-9}$ & 0.085 & 720 \\
\hline
\end{tabular}
\end{table}
$\kappa_{abs}$ is found to amount to 6.5 GHz for the WGM employed on the InGaP disk (TE, $p=3$), and to $\sim$ 117 GHz for the WGM (TE, $p=1$) employed on the AlGaAs disk. Based on this analysis, and knowing the structure thermoelastic properties, the photothermal force per photon $F_{pth}^1$ can be computed. It is listed in Table 1, together with the single photon optical force associated to radiation pressure and electrostriction. Once all these parameters evaluated, the threshold of optomechanical self-oscillation is obtained by equalling $\Gamma_{m,eff}$ to zero. $\Gamma_{m,eff}$ involves the thermal response time of the disk $\tau_{th}$, which can be obtained by FEM thermal simulations (see Appendix) or by fitting precisely the evolution of the mechanical linewidth $\Gamma_{m,eff}$ as a function of detuning, taking $\tau_{th}$ as an adjustable parameter (see Appendix). Both approaches give consistent outcomes, considering the uncertainty resulting from FEM thermal simulations. Indeed the pedestal dimensions and exact shape are estimated with a limited precision due to the SEM, which introduces an imprecision in the simulated thermal relaxation time. The values of $\tau_{th}$ reported in Table 1 were obtained from a fit of the mechanical linewidth by our optomechanical model. We notice that $\tau_{th}$ is smaller in the AlGaAs disk compared to InGaP. This originates from distinct pedestal geometries. The AlGaAs disk used here differs from that shown in Fig. 1(b) and has a pedestal radius $\sim$ \( 1.05\ \mu \)m, while that of the InGaP disk has a radius of \( 620\ \)nm. In Table 1 we observe that the photothermal force dominates the two other forces. From our modeling, it also appears that the photothermal force brings a large contribution to the optomechanical amplification of motion, even if radiation pressure and electrostriction do also participate (because it is a dynamical effect, the static force values reported in Table 1 are not the only parameters to matter for in the amplification regime). More generally, the relative importance of the three forces is strongly system-dependent. A smaller radius tends to increase the $g_{om}$ values and rapidly gives importance to radiation pressure and electrostriction. In parallel, a reduced absorption rate, obtained for example by ALD surface treatment \cite{Guha2017}, and an operation at low temperature that lowers thermal expansion, will strongly minimize photothermal forces. In consequence, the weight of optical forces at play (radiation pressure, electrostriction and photothermal) strongly depends on the detail of the resonator geometry and on operating conditions.  

\section{Conclusion}

The best optical and mechanical quality factor measured on optomechanical disk resonators of this work are listed in Table 2, and compared with previously published results on GaAs disk resonators. We see that the best photonic performances are obtained on GaAs disks with surfaces treated by ALD \cite{Guha2017}. However, the two new optomechanical materials reported here bear interesting potential. Thanks to the absence of two photon absorption at \( 1.5\ \mu \)m, the dissipative non-linearities and instabilities at high optical power are reduced compared to the case of GaAs disks. In the long run, this should allow for extremely large cooperativities to be supported. 
\begin{table}[h!]
\centering
\captionsetup{font=bf}
\captionsetup{labelsep=period}
\caption{Best Q$_{\text{opt}}$ and Q$_{\text{m}}$ (intrinsic values), along with the frequency of the 1$^{\text{st}}$ order RBM, measured on different disk resonators. Measurements are with an optical wavelength range 1500-1600 nm and at room temperature.}
\smallskip
\resizebox{\linewidth}{!}{
\begin{tabular}{c  c  c  c  c c}
\hline
 \textbf{Disk}
 &
 \begin{tabular}{c}
 \textbf{Radius} \\
 \textbf{(}$\boldsymbol{\mu}$\textbf{m)}
 \end{tabular}
 &
 \begin{tabular}{c}
 \textbf{Thickness} \\
 \textbf{(nm)}
 \end{tabular}
 &
 \textbf{Q$_{\text{opt}}$}
 &
 \begin{tabular}{c}
 $\boldsymbol{f_m}$ \\
 \textbf{(MHz)}
 \end{tabular}
 &
 \begin{tabular}{c}
 \textbf{Q$_{\text{m}}$} \\
 \textbf{in air at room T}
 \end{tabular} \\
\hline
\textbf{InGaP} & 3.25 & 200 & $\sim 1\times 10^5$ & 480.77 & $\sim 1500$ \\
\textbf{AlGaAs} & 2 & 150 & $\sim 5\times 10^4$ & 757.10 & $\sim 350$ \\
\textbf{GaAs} & 4.5 & 200 & $\sim 5\times 10^5$ & 314.5 & $\sim 3000$ \\
\textbf{GaAs with ALD} & 4.5 & 200 & $\sim 6\times 10^6$ & $\sim 314.5$ & $\sim 3000$ \\
\hline
\end{tabular}}
\end{table}

Of course, a lot of improvements are still required to reveal this potential in full. For instance, the fabricated InGaP disks deviate from a truly circular shape and their selective wet etching reveals some faceting. On the other hand, AlGaAs surface easily oxidizes in air, which is anticipated to introduce optical losses, and indeed we inferred the presence of optical absorption in AlGaAs disks (notably on the surfaces, but maybe in the bulk as well). Surface control techniques \cite{Guha2017} will hence be required to obtain the best out of these new resonators. Be they finally surpassing other platforms or not, InGaP and AlGaAs optomechanical resonators will certainly enrich our understanding of fundamental photon/phonon interactions in III-V heterostructures, and make a broader range of applications at reach for optomechanics.

\section*{Appendix}

\section*{Disk resonators on a mesa}

The disks are elevated over the substrate on a mesa. Figures 4(a) and (b) show InGaP disk resonators positioned on such mesa. The WGMs of the disk resonator are excited by evanescent coupling to a tapered fiber. The inset of 4(a) is a schematic representation of this technique.   

\begin{figure}[htbp]
\centering\includegraphics[width=\textwidth]{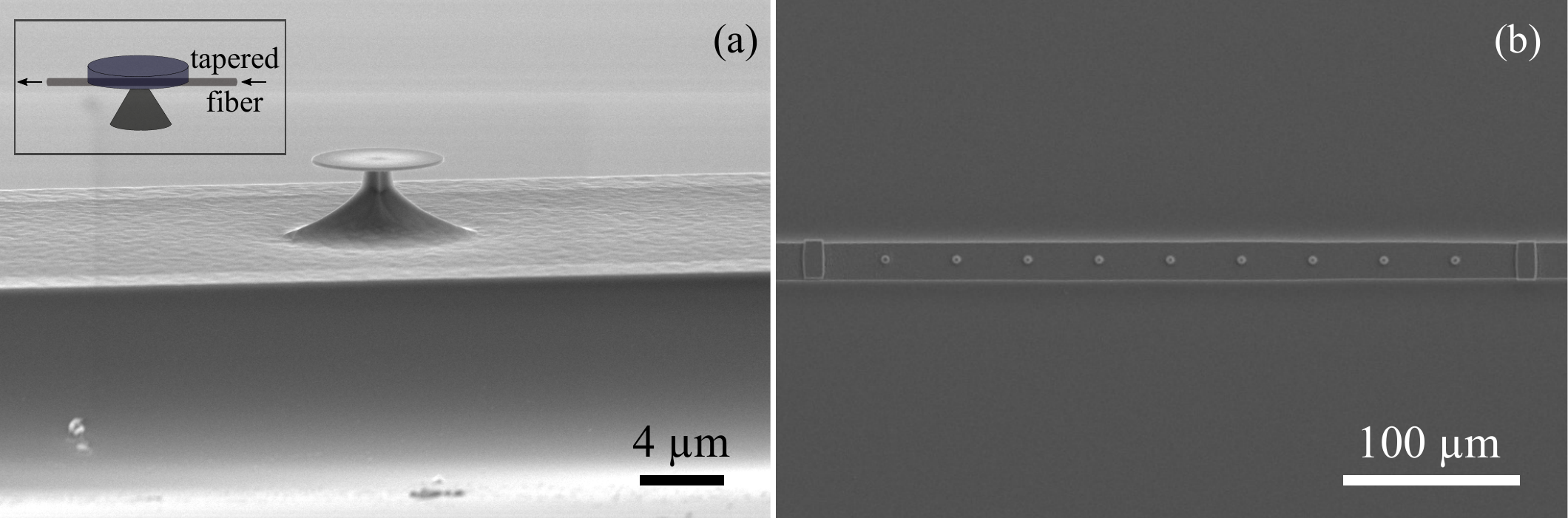}
\captionsetup{labelsep=period}
\caption{(a) An In$_{0.5}$Ga$_{0.5}$P disk on a mesa. Inset is a schematic representation of the evanescent coupling of light from a tapered fiber to a disk resonator. (b) Several In$_{0.5}$Ga$_{0.5}$P disks on a mesa. }
\end{figure}

\section*{Thermal and thermo-optic effects}

The disk thermal relaxation time $\tau_{th}$ is calculated using time-dependent FEM simulations of the heat transfer from the disk to the substrate. Figure 5(a) shows the steady-state temperature profile in an InGaP disk, for \( 1\ \)mW of power absorbed at the disk periphery. We observe that the temperature is almost uniform at the disk periphery, where the WGMs sit. The temperature of the disk in this zone as a function of time is shown in Fig. 5(b). $\tau_{th}$ is evaluated by fitting this plot with an exponential function. The AlGaAs disk investigated in Figs. 2 and 3 has a larger pedestal than its InGaP counterpart, which results in a faster thermal relaxation. Figure 5 shows the thermo-optic shift of a WGM resonance in the same AlGaAs disk. Figure 5(c) is the experimental measurement and (d) the result of the analytical model presented in \cite{Parrain2015} and adapted to the present case. The data/model agreement allows an evaluation of the residual linear optical absorption $\kappa_{abs}$ within the resonator, and is a first step to model the photothermal force. 

\begin{figure}[htbp]
\centering\includegraphics[width=\textwidth]{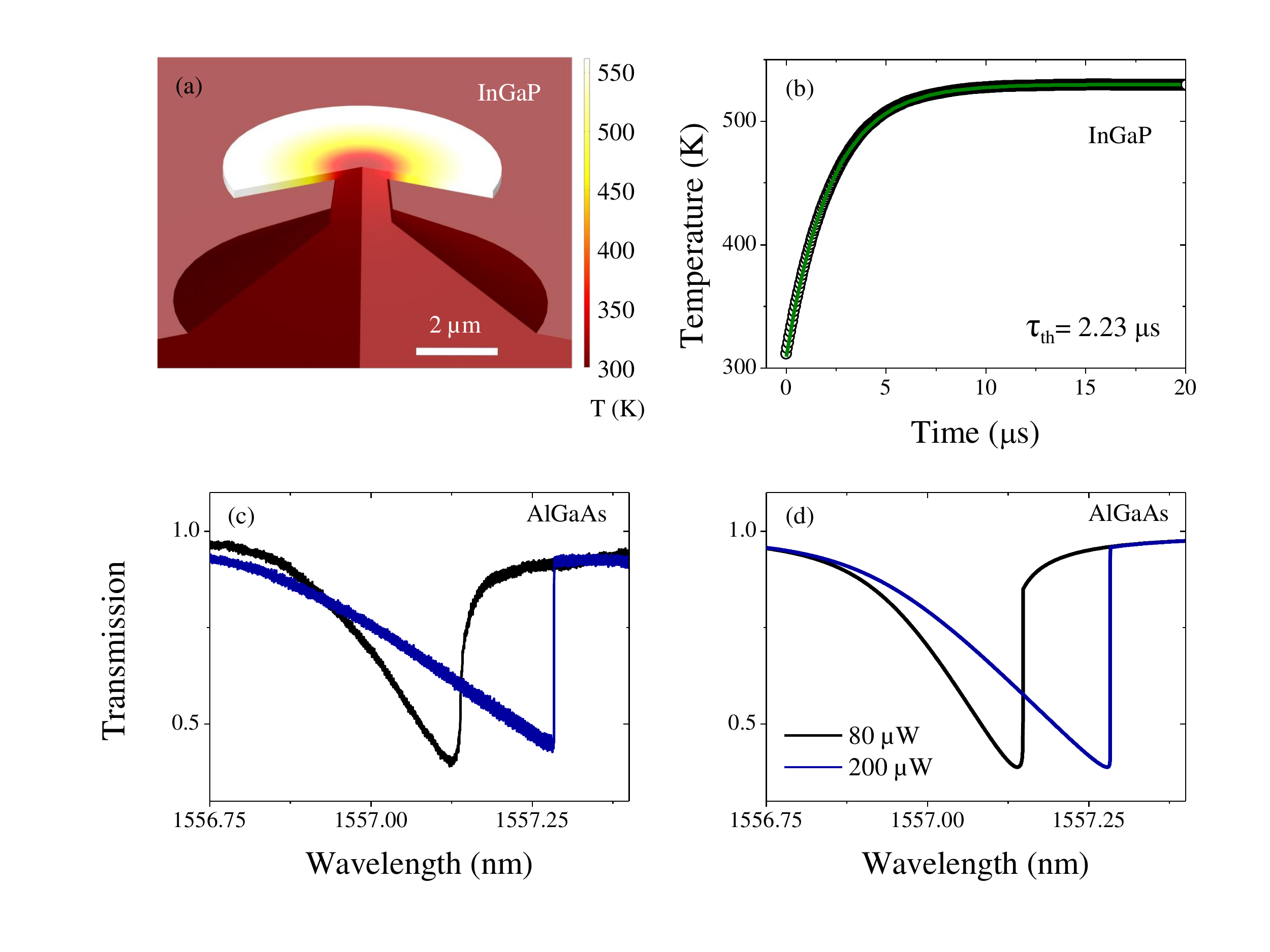}
\captionsetup{labelsep=period}
\caption{(a) Temperature profile (steady-state) in an InGaP disk, for \( 1\ \)mW absorbed power at the disk periphery. The colour bar indicates the temperature in K. (b) Temperature as a function of time. The green line corresponds to an exponential fit. (c,d) Thermo-optic shift of a WGM resonance in an AlGaAs disk. (c) Experimental measurements. (d) Results of our model \cite{Parrain2015}. The indicated power levels are measured at the fiber output.}
\end{figure}

\section*{Analytical optomechanical model for the self-oscillation}

The threshold of self-oscillation is obtained analytically by setting the effective damping ($\Gamma_{m,eff}$) of mechanical motion to zero. $\Gamma_{m,eff}$ is derived from the linearized equations of optomechanics including radiation pressure, electrostriction and photothermal forces. The three coupled differential equations involving optics, mechanics, and temperature of the disk read

\begin{equation}
\dot{a}(t) = -\frac{\kappa}{2}a(t)+i\left[ \Delta^b\omega + g_{om}x(t) + \frac{\omega_{cav}}{n}\frac{dn}{dt}\Delta T(t)\right] a(t) + \sqrt{\kappa_{ex}}a_{in}(t)
\end{equation}

\begin{equation}
m_{eff}\ddot{x}(t) + m_{eff}\Gamma_m\dot{x}(t) + m_{eff}\omega_m^2 x(t) =  F_{opt}(t)+F_{pth}(t)+F_{L}(t) 
\end{equation}

\begin{equation}
\frac{d\Delta T(t)}{dt} = -\frac{\Delta T(t)}{\tau_{th}}+\frac{\Gamma_{pth}|a(t)|^2}{\tau_{th}} 
\end{equation}

Here $a(t)$ is the complex optical field amplitude in the cavity normalized such that $|a(t)|^2$ gives the number of cavity photons, $\kappa=\kappa_{in}+\kappa_{ex}$ is the total decay rate of the optical energy stored in the cavity, $\Delta^b \omega=\omega_L - \omega_{cav}$ is the laser frequency detuning to the bare cavity, $x(t)$ is the mechanical displacement, $\omega_{cav}$ is the bare cavity resonance frequency, $\frac{dn}{dt}$ is the thermo-optic coefficient of the disk material, $\Delta T$ is the increase in disk temperature. In steady state $\Delta T=\Gamma_{pth}|a|^2$ with $\Gamma_{pth}=R_{th}\hbar\omega_L\kappa_{abs}$ where $R_{th}$ is a thermal resistance that links the temperature increase $\Delta T$ to the intra-cavity absorbed power. $R_{th}$ can be calculated using the above FEM thermal simulations in the steady-state regime. $\omega_m$ is the mechanical frequency, $m_{eff}$ is the effective mass of the mechanical mode, $\Gamma_m=\frac{\omega_m}{\text{Q}_\text{m}}$ is the mechanical decay rate. These coupled equations involve the non-dissipative optical forces $F_{opt}=F_{rp}+F_{el}$, the dissipative photothermal force $F_{pth}$ and the thermal Langevin force $F_{L}$. Linearizing these three equations and moving in the Fourier space, we derive an effective mechanical frequency $\omega_{m,eff}$ and an effective mechanical damping $\Gamma_{m,eff}$ for the mechanical system interacting with the optical cavity.
\begin{table}[h!]
\captionsetup{font=bf}
\captionsetup{labelsep=period}
\caption{Material properties of GaAs, In$_{\boldsymbol{0.5}}$Ga$_{\boldsymbol{0.5}}$P and Al$_{\boldsymbol{0.4}}$Ga$_{\boldsymbol{0.6}}$As disks \cite{Combrie2009, Mariani2014, Blakemore1982, Corte2000, Adachi1985, Ioffe_database, Haynes2012, Adachi1992, Gandomkar2011, Sokolov2017, Baker2013}.}
\smallskip
\resizebox{\linewidth}{!}{
\centering
\begin{tabular}{c c c c c}
\hline
\textbf{Property} & \textbf{unit}
&
\textbf{GaAs}
& 
\textbf{In$\boldsymbol{_{0.5}}$Ga$\boldsymbol{_{0.5}}$P}
&
\textbf{Al$\boldsymbol{_{0.4}}$Ga$\boldsymbol{_{0.6}}$As}
\\
\hline
\hline
\begin{tabular}{c}
\textbf{Poisson's} \\
\textbf{ratio (}$\boldsymbol{\nu}$\textbf{)}
\end{tabular}
& --- & 0.31 & 0.3345 & 0.35
\\ 
\hline
\begin{tabular}{c}
\textbf{Young's} \\
\textbf{modulus (}$\boldsymbol{E}$\textbf{)}
\end{tabular}
& GPa & 85.9 & 82.5 & 84.6
\\
\hline
\textbf{Density (}$\boldsymbol{\rho}$\textbf{)} & Kg/m$^3$ & 5317 & 4470 & 4696
\\
\hline
\begin{tabular}{c}
\textbf{Specific} \\
\textbf{heat (}$\boldsymbol{C}$\textbf{)}
\end{tabular}
& JKg$^{-1}$K$^{-1}$ & 327 & 371.2 & 378
\\
\hline
\begin{tabular}{c}
\textbf{Thermal} \\
\textbf{conductivity (}$\boldsymbol{\lambda}$\textbf{)}
\end{tabular}
& WK$^{-1}$m$^{-1}$ & 55 & 5.26 & 9.88
\\
\hline
\begin{tabular}{c}
\textbf{Thermal} \\
\textbf{expansion} \\ \textbf{coefficient (}$\boldsymbol{\alpha}$\textbf{)}
\end{tabular}
& K$^{-1}$ & 5.7$\times 10^{-6}$ & 5.3$\times 10^{-6}$ & 5.52$\times 10^{-6}$
\\
\hline
\begin{tabular}{c}
\textbf{Refractive index (}$\boldsymbol{n}$\textbf{)} \\
\textbf{at} $\boldsymbol{\lambda}$\textbf{= 1550 nm}\\
\textbf{and room T} 
\end{tabular}
& --- & 3.374 & 3.19 & 3.17
\\
\hline
\begin{tabular}{c}
\textbf{Thermo-optic}\\
\textbf{coefficient (}$\boldsymbol{\frac{\partial n}{\partial T}}$\textbf{)} \\
\textbf{at room T}
\end{tabular}
& K$^{-1}$ & 2.34$\times 10^{-4}$ & 2$\times 10^{-4}$ & 2.3$\times 10^{-4}$
\\
\hline
\begin{tabular}{c}
\textbf{TPA coefficient (}$\boldsymbol{\beta}$\textbf{) at}\\
$\boldsymbol{\lambda}$\textbf{= 1550 nm} 
\textbf{and room T} 
\end{tabular}
& cm/GW & $\sim 15$ & $\sim 0$ & $\sim 0$
\\
\hline
\begin{tabular}{c}
\textbf{Photoelastic parameters}\\
$\boldsymbol{p_{11}}$ \\
$\boldsymbol{p_{12}}$ \\
$\boldsymbol{p_{44}}$
\end{tabular}
& --- &
\begin{tabular}{c}
\\
-0.165 \\
-0.140 \\
-0.172
\end{tabular} 
& 
\begin{tabular}{c}
\\
-0.151 \\
-0.082 \\
-0.074
\end{tabular} 
& 
\begin{tabular}{c}
\\
-0.165 \\
-0.090 \\
-0.088
\end{tabular} 
\\
\hline
\hline
\end{tabular}} 
\label{table_comparative_study_mat_properties}
\end{table}
$\Gamma_{m,eff}$ reads
\begin{multline}
\Gamma_{m,eff}=  \Gamma_m\left[ 1+ \frac{|<a>|^2 g_{om}^2\omega_m}{m_{eff}\omega_m\Gamma_m}\left\lbrace \frac{\frac{\kappa}{2}}{(\Delta\omega+\omega_m)^2+\frac{\kappa^2}{4}} - \frac{\frac{\kappa}{2}}{(\Delta\omega-\omega_m)^2+\frac{\kappa^2}{4}} \right\rbrace \right. \\
 \left. + \frac{|<a>|^2 g_{om}\omega_m}{\hbar m_{eff}\omega_m\Gamma_m} \frac{F_{pth}^1}{1+\omega_m^2\tau_{th}^2}\left\lbrace  \frac{(\Delta\omega+\omega_m)\omega_m\tau_{th}-\frac{\kappa}{2}}{(\Delta\omega+\omega_m)^2+\frac{\kappa^2}{4}}+ \frac{(\Delta\omega-\omega_m)\omega\tau_{th}+\frac{\kappa}{2}}{(\Delta\omega-\omega_m)^2+\frac{\kappa^2}{4}} \right\rbrace \right] 
\end{multline}
where the detuning to the shifted cavity frequency is $\Delta\omega=\Delta^b\omega + g_{om}x_{eq} + \frac{\omega_{cav}}{n}\frac{dn}{dt}\Delta T_{eq}$ with $\Delta T_{eq}$ the mean temperature increase around which the thermal dynamics is linearized, $|<a>|^2$ the mean photon number in the cavity and $x_{eq}$ is the mean displacement around which the linearization is carried out. $g_{om}=g_{om}^{geo}+g_{om}^{pe}$. FEM computed values of $g_{om}^{geo}$ and $g_{om}^{pe}$ are 182 GHz/nm and 254 GHz/nm for the considered $p=3$ TE-WGM of the InGaP disk and 237 GHz/nm and 324 GHz/nm for the considered $p=1$ TE-WGM of the AlGaAs disk respectively. The material properties of In$_{0.5}$Ga$_{0.5}$P and Al$_{0.4}$Ga$_{0.6}$As are listed in Table 3, together with GaAs for comparison. A noticeable difference is the large thermal conductivity of GaAs compared to the two other materials. With all these parameters and the evaluation of the photothermal force, one can fit the systematic evolution of $\Gamma_{m,eff}$ as a function of detuning, as shown in Fig. 6 for the two disk resonators studied in Fig. 3. This can be done either by using the FEM-simulated value of $\tau_{th}$, or by letting $\tau_{th}$ as an adjustable parameter to exactly fit the value of self-oscillation threshold. Both approaches are consistent at the level of precision reached here. 

\begin{figure}[htbp]
\centering\includegraphics[width=\textwidth]{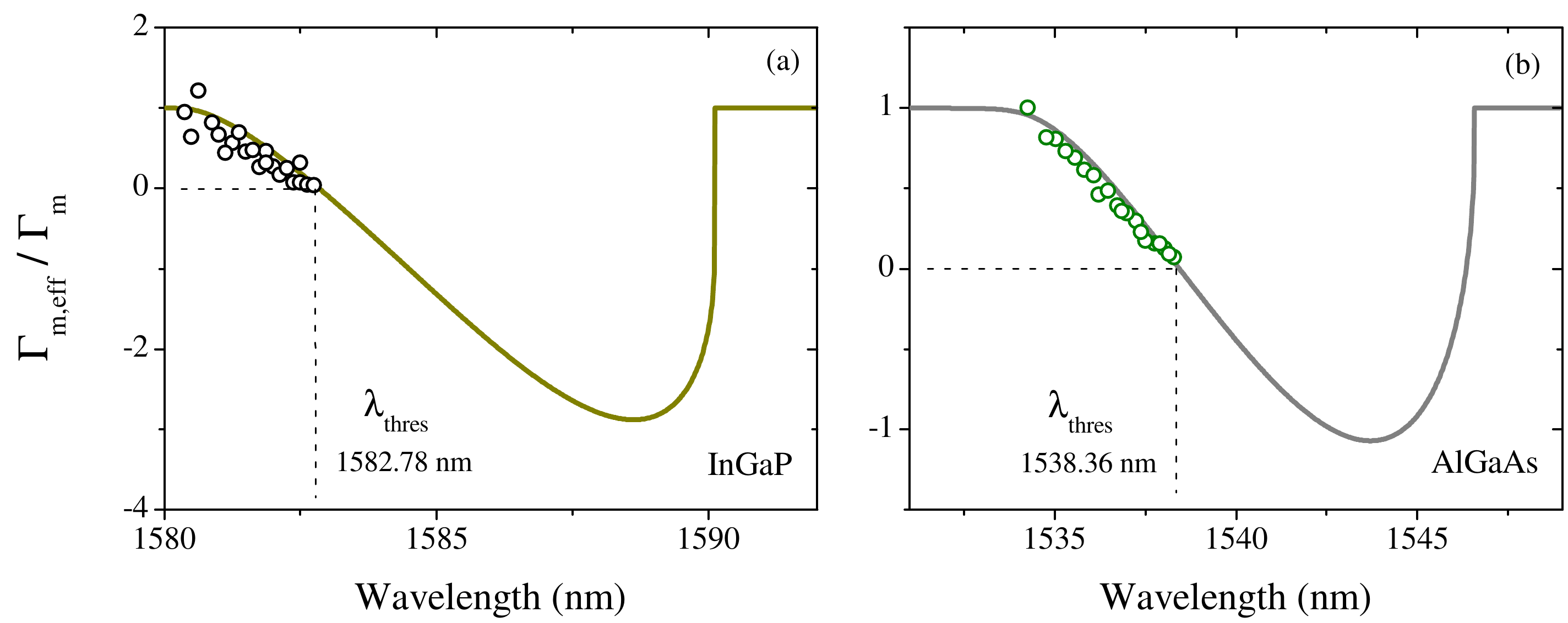}
\captionsetup{labelsep=period}
\caption{Mechanical linewidth as function of the laser wavelength, extracted from the self-oscillation spectra series of Fig. 3 for both the InGaP (a) and AlGaAs (b) disk resonator. The open symbols are experimental data, while the solid line is a fit by our optomechanical model, with a thermal time $\tau_{th}$=1.8 (0.085) $\mu$s for the InGaP (AlGaAs) resonator. The dashed lines indicate the self-oscillation threshold.}
\end{figure}

\section*{Funding}

French Agence National de la Recherche (ANR) through the QDOM project, European Research Council (ERC) through the GANOMS project (306664).

\section*{Acknowledgment}

Sylvain Combri\'{e} acknowledges the support of the European Commission through the H2020-FETPROACT-2016 project n. 732894 \textbackslash HOT.


\end{document}